# COGNITIVE RADIO ENGINE MODEL UTILIZING SOFT FUSION BASED GENETIC ALGORITHM FOR COOPERATIVE SPECTRUM OPTIMIZATION

Md. Kamal Hossain, Ayman Abd El-Saleh

Faculty of Engineering, Multimedia University, Cyberjaya Campus,
Malaysia
md.kamal.hossain@hotmail.com

*ABSTRACT*

*Cognitive radio (CR) is to detect the presence of primary users (PUs) reliably in order to reduce the interference to licensed communications. Genetic algorithms (GAs) are well suited for CR optimization problems to increase efficiency of bandwidth utilization by manipulating its unused portions of the apparent spectrum. In this paper, a binary genetic algorithm (BGA)-based soft fusion (SF) scheme for cooperative spectrum sensing in cognitive radio network is proposed to improve detection performance and bandwidth utilization. The BGA-based optimization method is implemented at the fusion centre of a linear SF scheme to optimize the weighting coefficients vector to maximize global probability of detection performance. Simulation results and analyses confirm that the proposed scheme meets real time requirements of cognitive radio spectrum sensing and it outperforms conventional natural deflection coefficient- (NDC-), modified deflection coefficient- (MDC-), maximal ratio combining- (MRC-) and equal gain combining- (EGC-) based SDF schemes as well as the OR-rule based hard decision fusion (HDF). The propose BGA scheme also converges fast and achieves the optimum performance, which means that BGA-based method is efficient and quite stable also.*

## 1. INTRODUCTION

In wireless communication system, electromagnetic spectrum is in scarcity considering the availability of resources. With the burgeoning of technology, the demand for the spectrum is also increasing insistently which results scarcity in spectrum-availability. Previous assumption about the crisis of spectrum availability resulted misconception. It is resolved by the Federal Communications Commission (FCC) [1] that underutilization of the licensed spectrum bands either temporally or spatially is the principal reason for spectrum scarcity. In order to fully utilize the spectrum resources, CR [2], i.e. radio systems with adaptive intelligence, is fascinating researchers and developers to overtake spectrum congestion bottlenecks. Cognitive radio takes advantage of the rapidly increasing complexity of radio equipment. It has been considered as the key enabling technology for current spectrum scarcity problem arising because of the necessity for secure and robust communications and is becoming more apparent every day. Wireless services are becoming largely ubiquitous, despite its expensive implementation. The CR users are considered as secondary users (SUs) in Cognitive Radio Network (CRN).

The main idea of CR is to periodically monitor the radio spectrum, detect the occupancy and then opportunistically use spectrum holes with minimal interference with primary users (PUs). In order to detect the PU signal with unknown location, structure and strength, energy detection exhibits simplicity and serves as the optimal spectrum sensing scheme. However, owing to hidden terminal problem and shadowing effect, the PU signal packets might not be received by CR receiver within the sharp sensing interval and thus the sensing performance will be rendered fragile at a particular geographical location [3]. This amplifies potential interference caused to the PU. In other words, energy detection is extremely vulnerable to the channel effects such as multipath fading and noise-power fluctuations that supposed to be avoided. In [4], [5], cooperative spectrum sensing was used to overcome these drawbacks and vacate the band immediately if PUs' presence has been detected.

The decision on the presence of PU is achieved by combining all individual decisions of local SUs at a central Fusion Centre (FC) using various fusion schemes [6], [7]. These schemes can be classified as hard decision fusion (HDF) [4], [8], [9], soft decision fusion (SDF) [10], [11] or softened hard decision fusion (SHDF) [11], [12]. In HDF, the local sensors, or SUs, make their own judgments on the presence of a PU and their corresponding resultant 1-bit decisions are sent to the FC for fusion. Apparently the SDF schemes want the local sensors to report their measurements as raw data to the FC where data will be fused to construct a final decision on the presence of PUs. In [12], [13] SDF-based schemes have shown better detection performance than HDF schemes. In [11], cooperative spectrum sensing was proposed and NDC, MDC, MRC, EGC, OR-Rule based methods were used to find the optimal weighting vector for all possible cognitive radios. In this paper, a BGA-based soft decision fusion (SDF) scheme for cooperative spectrum sensing in cognitive radio network is proposed to enhance the detection performance.

A genetic algorithm (GA) is a search heuristic that mimics the process of natural evolution. GAs are well equipped with many tools to reduce computational complexity and produce a diverse set of solutions which can be implemented on semiconductor devices and enable rapid integration with wireless technologies. Genetic algorithm is a kind of self-adaptive global searching optimization algorithm. It is population-based in which each individual is evolved in parallel and the optimal individual is preserved and obtainable from the last population. The contributions of this paper are listed as follows:

(a) In order to improve detection performance of cooperative spectrum sensing in Cognitive Radio Network (CRN), PU detection problem has been reformulated using BGA.

(b) Then, the proposed method will be compared with conventional NDC, MDC, MRC and EGC based SDF schemes as well as the OR-rule based hard decision fusion (HDF) to verify the supremacy of the proposed method.

The rest of this paper is structured as follows: Section 2 presents some related research background. Section 3 briefly explains system model, while Section 4 shows comparisons and results of performance of Genetic Algorithm over other methods. Finally, Section 5 concludes the paper and provides future works.

## 2. RELATED WORKS

CR Technology has been evolved as key enabling technology to the problem of spectrum underutilization. CR designed to allow unlicensed or SUs to access spectrum bands which has been allocated to PUs when interference to PU remains below a given threshold. The main inspiration behind this CR technology is the new spectrum license initiated by the FCC, which will be more flexible to allow unlicensed or SUs to access the spectrum as long as the licensed or PUs are not interfered. That's why there is an increased interest of researchers in this technology in academia, industry and engineers in the wireless industry and also from spectrum policy makers.

In [5], [12], the authors proposed an optimal weighting scheme for cooperative spectrum sensing in cognitive radio networks, under the constraint of equal probabilities of false alarm and miss detection. Multiple cooperative SUs simply serve as relay nodes in the network to provide space diversity for spectrum sensing. An optimal soft fusion scheme and a double-threshold strategy were proposed to investigate the overall spectrum sensing performance in soft fusion and hard fusion, respectively. From their observations and analysis proposed hierarchical cooperative spectrum sensing schemes can achieve significant improvements in spectrum sensing performance.

In [13], GA-based weighted collaborative spectrum sensing strategy was proposed to reduce the effects of channel and enhance spectrum sensing performance. Authors proposed optimum spectrum sensing framework is based on a model that is realistic and also takes into account both channels, that is, channel between PU and SUs as well as the reporting channels. It was shown in this paper that imperfect reporting channel and different SU SNR values have direct impact on the performance of CSS. SUs transmit their soft decisions to the fusion centre and a global decision is made at the fusion centre which is based on a weighted combination of the local test statistics from individual SUs. The weight of each SU is indicative of its contribution to the final decision making.

Finally in [14], the authors addressed SDF-based scheme to exploit the advantage of optimum detection performance of SDF. The SDF schemes had been implemented using weighting coefficients vector based on NDC, MDC, MRC and EGC based SDF schemes as well as the OR-rule based hard decision fusion (HDF). These SDF schemes were implemented within the cluster. The 1-bit PU-availability decisions of several cluster then forwarded to a common receiver (BS) at which an OR-Rule HDF scheme used to come out with a global single decision on the presence of a PU. Authors' analysis concerned SDF performs better than HDF scheme.

In our approach, we are going to implement BGA-based soft decision fusion (SDF) scheme for cooperative spectrum sensing in cognitive radio network.

### 3. SYSTEM MODEL

In Cognitive Radio Networks (CRNs), the detection performance might be vulnerable when the sensing decisions forwarded to a fusion centre through fading channels. A SDF-based cooperative spectrum sensing is used to improve detection reliability. Figure 1 shows a deployment of CRN using cooperative spectrum sensing.

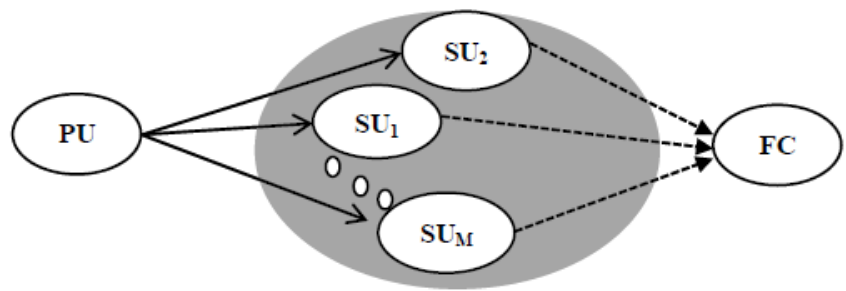

Figure 1: SDF-base Cooperative spectrum sensing in a CRN

As seen in Figure 1, $M$ SUs are relaying their individual statistical measurements of PU availability to a common FC. The FC functions as a decision centre which manages the CR network as well all associated SUs. The use of a weighting vector in the linear soft fusion helps to eliminate the need for finding optimal thresholds for individual SU. Figure 1 also presents two progressive links, namely, primary user-secondary user (PU-SU) link and secondary user-fusion centre (SU-FC) link. The main operations carried out on these two links are spectrum sensing and SDF, respectively.

### 3.1 Characterization of Primary User-Secondary User (PU-SU) link

SUs in the network are grouped into multiple clusters by some upper layer distributed clustering algorithms [15]. The SU performs local spectrum sensing individually to detect PU's presence. The sensing technique is formulated as binary hypothesis test.

When PU is absent $\Rightarrow H_o: X_i[n] = W_i[n]$ (1)

When PU is present $\Rightarrow H_1: X_i[n] = g_i S[n] + W_i[n]$ (2)

where $X_i[n]$ is the received sampled signal at the $i^{th}$ SU receiver, $n = 1, 2, \ldots, K$, where $K$ is the number of samples of the received signal and it is defined as $K = 2T_sB$ where $T_s$ is the sensing time, $i = 1, 2, \ldots, M$, where $M$ is the number of cooperative SUs, $g_i$ is the sensing channel gain between the PU and the $i^{th}$ SU which accommodates for any channel effects such as multipath fading, shadowing, and propagation path loss, $S[n]$ is the PU transmitted signal which is assumed to be independent and identically distributed (i.i.d.) Gaussian random process with zero mean and variance $\sigma_S^2$, i.e., $S[n] \sim N(0, \sigma_S^2)$, and $W_i[n]$ is the $i^{th}$ sensing channel noise which is assumed to be additive white Gaussian with zero mean and variance $\sigma_{Wi}^2$ experiencing i.i.d. fading effects, i.e., $W_i[n] \sim N(0, \sigma_{Wi}^2)$. All these variances are collected into the vector $\vec{\sigma}_W = [\sigma_{W1}^2, \sigma_{W2}^2, \ldots, \sigma_{WM}^2]^T$ and the sampled signals received at the $M$ SUs are collected into the vector $\vec{X} = [X_1, X_2, \ldots, X_M]^T$. The channel gains of the PU-SU and SU-FC links, $g_i$ and $h_i$, respectively, are assumed to be constant over each sensing period; this can be justified by the slow-fading nature over these links where the delay requirement is short compared to the channel coherence time considered as quasi-static scenario [16]. A detailed system model simplified from [14] for the SDF-based cooperative spectrum sensing is shown in Figure 2.

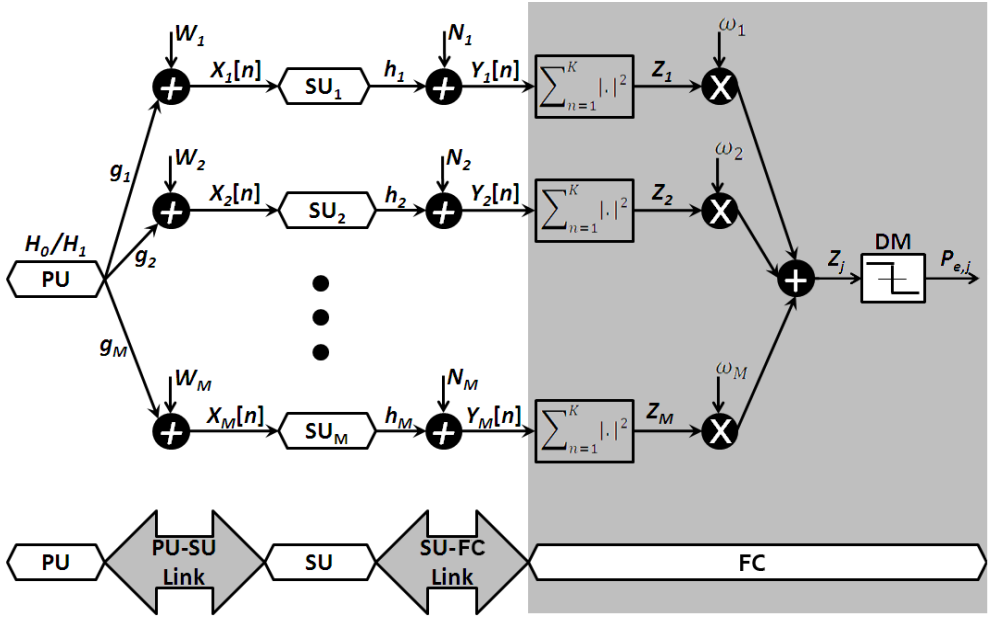

Figure 2. Detailed system model of SDF-based cooperative spectrum sensing

### 3.2. Characterization of Secondary User- Fusion Centre (SU-FC) link

In [14], the SDF process is initialized by notifying the *M* SUs to relay their individual measurements of PUs' signal, *X*, FC through a dedicated control channel in an orthogonal manner. In the paper the justification of using amplify-and-forward (AAF) instead of the less complexity decode-and-forward (DAF) scheme is mentioned to its ability to increase the detection performance by employing some signal processing techniques at the FC. The channel noises $\{N_i\}$ of the SU-FC links are expected to be zero mean and spatially uncorrelated additive white Gaussian with variances $\{\delta_i^2\}$ which has been composed into the vector $\vec{\delta} = [\delta_1^2, \delta_2^2, ..., \delta_m^2]^T$. Then, the signal received by corresponding *FC* from the $i^{th}$ SU will be

$$Y_i[n] = \sqrt{P_{Ri}}\, h_i\, X_i[n] + N_i[n] \qquad (3)$$

where $P_{Ri}$ = the transmit power of the $i^{th}$ relay $h_i$ =is the amplitude channel gain of the SU-FC link.

The use of AWGN model here is justified by the slow-changing nature of the channels between the *M* SUs and their corresponding FC. Now, by considering the two hypotheses in (1) and (2), the received signal at FC can be expressed as

$$Y_i[n/H_o] = \sqrt{P_{Ri}}\, h_i\, W_i[n] + N_i[n] = u_{0i}[n] \qquad (4)$$

$$Y_i[n/H_1] = \sqrt{P_{Ri}}\, h_i\, g_i\, S[n] + \sqrt{P_{Ri}}\, h_i\, W_i[n] + N_i[n] = \sqrt{P_{Ri}}\, h_i\, g_i\, S[n] + u_{0i}[n] \qquad (5)$$

whose statistical properties are $Y_i[n/H_o] \sim \mathcal{N}(0, \sigma_{0,i}^2) \sim \mathcal{N}(0, P_{Ri}|h_i|^2 \sigma_{Wi}^2 + \delta_i^2)$ and $Y_i[n/H_1] \sim \mathcal{N}(0, \sigma_{1,i}^2) \sim \mathcal{N}(0, P_{Ri}|g_i|^2|h_i|^2 \sigma_S^2 + \sigma_{0,i}^2)$.

In a Matrix form, the received signals at the FC through the control channel under $H_0$ and $H_1$, respectively, can be written as

$$Y[n|H_0] = \begin{bmatrix} \sqrt{P_{R1}}h_1 & 0 & \cdots & 0 \\ 0 & \sqrt{P_{R2}}h_2 & \cdots & 0 \\ \vdots & \vdots & \ddots & \vdots \\ 0 & 0 & 0 & \sqrt{P_{RM}}h_M \end{bmatrix} \times \begin{bmatrix} W_1[n] \\ \vdots \\ W_M[n] \end{bmatrix} + \begin{bmatrix} N_1[n] \\ \vdots \\ N_M[n] \end{bmatrix} \qquad (6)$$

$$Y[n|H_1] = \begin{bmatrix} \sqrt{P_{R1}}g_1 h_1 & 0 & \cdots & 0 \\ 0 & \sqrt{P_{R2}}g_2 h_2 & \cdots & 0 \\ \vdots & \vdots & \ddots & \vdots \\ 0 & 0 & 0 & \sqrt{P_{RM}}g_M h_M \end{bmatrix} \times \begin{bmatrix} S_1[n] \\ \vdots \\ S_M[n] \end{bmatrix} + \begin{bmatrix} u_{01}[n] \\ \vdots \\ u_{0M}[n] \end{bmatrix} \qquad (7)$$

At the corresponding FC of *M* SUs, each received sequence $Y_i[n]$ will be individually averaged and squared using a separate energy detector to estimate its own energy as shown in Figure 2. Thus, the estimated energy collected by the $i^{th}$ SU all the way to the FC can be expressed as

$$Z_i = \sum_{n=0}^{K-1} |Y_i[n]|^2 \; ; \qquad i = 1, 2, \ldots, M \tag{8}$$

By denoting $\{Z_{o,i}\} = \{Z_i|H_0\}$ and $\{Z_{1,i}\} = \{Z_i|H_1\}$, the two sets of test statistics can be written as $\vec{Z}_0 = [Z_{0,1}, Z_{0,2} \ldots Z_{0,M}]^T$ and $\vec{Z}_1 = [Z_{1,1}, Z_{1,2} \ldots Z_{1,M}]^T$. Here we have considered that for a large number of samples, the central limit theorem (CLT) approximates each test statistic into the vectors $\vec{Z}_0$ and $\vec{Z}_1$. The normally distributed with mean and variance given can be expressed by

$$E(Z_i | H_0) = K\sigma_{0,i}^2 = K(P_{Ri} |h_i|^2 \sigma_{Wi}^2 + \delta_i^2) = \mu_{0,i} \tag{9}$$

$$\text{var}(Z_i | H_0) = 2K\sigma_{0,i}^4 = 2K(P_{Ri} |h_i|^2 \sigma_{Wi}^2 + \delta_i^2)^2 \tag{10}$$

$$E(Z_i | H_1) = K\sigma_{1,i}^2 = K(P_{Ri} |g_i|^2 |h_i|^2 \sigma_S^2 + \sigma_{0,i}^2) = \mu_{1,i} \tag{11}$$

$$\text{var}(Z_i | H_1) = 2K\sigma_{1,i}^4 = 2K(P_{Ri} |g_i|^2 |h_i|^2 \sigma_S^2 + \sigma_{0,i}^2)^2 \tag{12}$$

where $\vec{\mu}_0 = [\mu_{0,1}, \mu_{0,2}, \ldots, \mu_{0,M}]^T$ and $\vec{\mu}_1 = [\mu_{1,1}, \mu_{1,2}, \ldots, \mu_{1,M}]^T$.

Let us assume here that $\theta_i = K P_{Ri} |g_i|^2 |h_i|^2 \sigma_S^2$ and $\vec{\theta} = [\theta_1, \theta_2, \ldots, \theta_M]^T$, then conclude that $\mu_{1,i} = \mu_{0,i} + \theta_i$ or $\vec{\mu}_1 = \vec{\mu}_0 + \vec{\theta}$.

Next, from Figure 4.2 that all the individual test statistics $\{Z_i\}$ are multiplied by weighting coefficient vector $\omega_i$ and used to linearly formulate the resultant test statistic of the FC, **Z**, which can be expressed as

$$\mathbf{Z} = \sum_{i=1}^{M} \omega_i Z_i = \vec{\omega}^T \vec{Z} \tag{13}$$

where the weighting coefficients vector $\vec{\omega} = [\omega_1, \omega_2, \ldots, \omega_M]^T$; $\omega_i \geq 0$ satisfying the condition; $\|\vec{\omega}\| = 1$ which is used to optimize the detection performance. Different weight settings will be addressed later. Since $\{Z_i\}$ are all normal random variables, their linear combination, which represent FC test statistic **Z**, will also be normally distributed with statistics given by

$$E(Z | H_0) = \sum_{i=1}^{M} \omega_i K(P_{Ri} |h_i|^2 \sigma_{Wi}^2 + \delta_i^2) = \sum_{i=1}^{M} \omega_i K\sigma_{0,i}^2 = \sum_{i=1}^{M} \omega_i \mu_{0,i} = \vec{\omega}^T \vec{\mu}_0 \tag{14}$$

$$E(Z | H_1) = \sum_{i=1}^{M} \omega_i K(P_{Ri} |g_i|^2 |h_i|^2 \sigma_S^2 + \sigma_{0,i}^2) = \sum_{i=1}^{M} \omega_i K\sigma_{1,i}^2 = \sum_{i=1}^{M} \omega_i \mu_{1,i} = \vec{\omega}^T \vec{\mu}_1 \tag{15}$$

$$\text{var}(Z | H_0) = \sum_{i=1}^{M} 2\omega_i^2 K(P_{Ri} |h_i|^2 \sigma_{Wi}^2 + \delta_i^2)^2 = \vec{\omega}^T \sum\nolimits_{H_0} \vec{\omega} \tag{16}$$

$$\text{var}(Z | H_1) = \sum_{i=1}^{M} 2\omega_i^2 K(P_{Ri} |g_i|^2 |h_i|^2 \sigma_S^2 + \sigma_{0,i}^2)^2 = \vec{\omega}^T \sum\nolimits_{H_1} \vec{\omega} \tag{17}$$

where the covariance matrices are $\sum\nolimits_{H_0} = 2K\sigma_{0,i}^4$ and $\sum\nolimits_{H_1} = 2K(P_{Ri} |g_i|^2 |h_i|^2 \sigma_S^2 + \sigma_{0,i}^2)^2$.

Considering that the global threshold at the FC $\beta$, the likelihood ratio is $Z \underset{H_0}{\overset{H_1}{\gtrless}} \beta$. As such, the overall probability of detection, $P_d$, and probability of false alarm, $P_f$, for the $M$ SUs of FC can be written as

$$P_f = P(Z > \beta \mid H_0) = Q\left(\frac{\beta - E(Z \mid H_0)}{\sqrt{\text{var}(Z \mid H_0)}}\right) = Q\left(\frac{\beta - \vec{\omega}^T \vec{\mu}_0}{\sqrt{\vec{\omega}^T \sum_{H_0} \vec{\omega}}}\right) \tag{18}$$

$$P_d = P(Z > \beta \mid H_1) = Q\left(\frac{\beta - E(Z \mid H_1)}{\sqrt{\text{var}(Z \mid H_1)}}\right) = Q\left(\frac{\beta - \vec{\omega}^T \vec{\mu}_1}{\sqrt{\vec{\omega}^T \sum_{H_1} \vec{\omega}}}\right) \tag{19}$$

In CRNs, the probabilities of false alarm and detection have unique indications. Specifically, $(1-P_d)$ measures the probability of interference from SUs on the PUs. On the other hand, $P_f$ determines an upper bound on the spectrum efficiency, where a large $P_f$ usually results in low spectrum utilization. This is because the SU is allowed to perform transmissions if and only if the PU is undetected under either $H_0$ or $H_1$. If the PU is undetected under either $H_0$ or $H_1$, Maximizing $P_d$ by controlling the weighting vector while meeting a certain requirement on the $P_f$ and vice versa. Then, for a given $P_f$, $P_d$ can be written as

$$P_d = Q\left(\frac{Q^{-1}(\overline{P}_f)\sqrt{\vec{\omega}^T \sum_{H_0} \vec{\omega}} - \vec{\omega}^T \theta}{\sqrt{\vec{\omega}^T \sum_{H_1} \vec{\omega}}}\right) \tag{20}$$

Similarly, for a given $P_d$, $P_f$ can be expressed as

$$P_f = Q\left(\frac{Q^{-1}(\overline{P}_d)\sqrt{\vec{\omega}^T \sum_{H_1} \vec{\omega}} + \vec{\omega}^T \theta}{\sqrt{\vec{\omega}^T \sum_{H_0} \vec{\omega}}}\right) \tag{21}$$

### 3.3. Conventional SDF-based cooperative spectrum sensing schemes

Here some conventional SDF optimization schemes for weighting vector setting at CHs are discussed as in [14]. These are NDC, MDC, MRC and EGC . The EGC scheme is an existing weighting scheme that is similar to the one used in systems with multiple receive antennas. . The individual weights assigned to the $M$ SUs signals at the FC in (18) and (19) are all equal and expressed by

$$\omega_i = \sqrt{1/M} \tag{22}$$

In MRC, the weight coefficient assigned for a particular SU signal at a FC represents its contribution to the overall decision made. Thus, if a SU has a high PU signal-to-noise ratio (SNR) at its receiver that may lead to a correct detection on its own, it should be assigned a larger weighting coefficient. For those SUs experiencing deep fading or shadowing, their weights are decreased in order to reduce their negative contribution to the final decision. By maintaining $\|\omega\| = 1$, individual weight for the $i^{th}$ SU's measurement for MRC as follows

$$\omega_i = \sqrt{\frac{SNR_i}{SNR_T}} \tag{23}$$

where $i = 1, 2, \ldots, N$ and $SNR_i$ is the signal-to-noise ratio at the CH receiver estimated for the $i^{th}$

SU.

The deflection coefficient (DC) is a measure of the detection performance as it is formulated based on the distance between the centers of $H_0$ and $H_1$. The DC based weight setting scheme can be realized by maximizing the normal DC or the modified DC as shown below.

NDC provides a good measure of the detection performance because the $\Sigma_{H_0}$ covariance matrix under hypothesis $H_0$ is used to characterize the variance-normalized distance between the centers of the two conditional PDFs of $Z_j$ under $H_0$ and $H_1$. To ensure $\|\vec{\omega}\|=1$ and normalizing each weighting co-efficient, we obtain the optimal weighting vector

$$\vec{\omega}^*_{opt,NDC} = \vec{\omega}_{opt,NDC} / \|\vec{\omega}_{opt,NDC}\| = \Sigma_{H0}^{-1}\vec{\theta} \qquad (24)$$

The maximization of MDC in order to find the optimal weights setting for the SDF at the CH. The MDC can be defined to ensure $\|\vec{\omega}\|=1$ and normalizing each weighting co-efficient, we obtain the optimal weighting vector

$$\vec{\omega}^*_{opt,MDC} = \vec{\omega}_{opt,MDC} / \|\vec{\omega}_{opt,MDC}\| = \Sigma_{H1}^{-1}\vec{\theta} \qquad (25)$$

.

## 4. PROPOSED SOLUTION

### 4.1 Proposed BGA based cooperative spectrum sensing

Genetic Algorithm (GA) is classified as an evolutionary algorithm that is a stochastic search method mimics natural evolution. GA is a kind of self-adaptive global searching optimization algorithm. It has been used to solve difficult problems like, Non deterministic problems and machine learning as well as also for simple programs like evolution of pictures and music. The main advantage of GAs over the other methods is their parallelism. GAs travel in a search space that uses more individuals for the decision-making and hence are less likely to get stuck in a local extreme like the other available decision-making techniques. It is a population-based in which each individual is evolved in parallel and the optimal individual is preserved and obtainable from the last set of population.

In general the genetic algorithm (GA) mechanism starts with randomly generating a set of chromosomes. These chromosomes constitute a population (*pops*). As the genetic algorithm models natural processes such as selection, crossover and mutation, it performs on a population of individuals instead of a single individual. A random population of chromosomes will be initialized and then will be evaluated by fitness functions of a particular problem. It will then check for optimization criterion defined by the engineer and will generate a new population from the previous populations if termination criterion is not met. These new individuals are selected according to their fitness values [17]. The chromosomes which are considered fit will be selected as parents and will undergo mating with crossover and mutation for better production. These new offspring will then be evaluated and becomes parents to the new generation if termination criterion is not satisfied. And this cycle will be looped until a certain criterion is met, where each iteration is called a generation.

In our proposed BGA method, an initial population of *pops* possible solutions is generated randomly and each individual is normalized to satisfy the constraints [3]. Our goal is to find the optimal set of weighting vector values to maximize detection performance. When it reaches the predefined maximum generation, BGA is terminated and the weighted vector values that makes highest fitness is considered as the best solution. Let us assume that we have $M$ SUs and $Z_1$, $Z_2$…$Z_M$ are the soft decisions of $SU_1$, $SU_2$….$SU_M$ on the presence of PUs, and $\vec{\omega}_j$ is the weighting vector of the $j^{th}$ individual that consists of $\omega_1, \omega_2, \ldots \omega_M$. Thus, the fitness value for the $j^{th}$ individual is defined as

$$f_j = P_d(\vec{\omega}_j) \quad \text{where } ||\vec{\omega}_j|| = 1 \tag{26}$$

The main operations of the proposed BGA are selection, crossover, and mutation. For selection, the idea is to choose the best chromosomes for reproduction through crossover and mutation. Larger the fitness value better the solution obtained. In this paper we use "Roulette Wheel selection" method. The probability of selecting the $j^{th}$ individual or chromosome, $p_j$, can be written as

$$p_j = \frac{f_j}{\sum_{j-1}^{pops} f_j} \tag{27}$$

The chromosomes with maximum probability value will be transferred to next generation through elitism operation. After selection process is done, the next step is crossover. The crossover starts with pairing to produce new offspring. We use a uniform random number generator to select the row numbers of chromosomes as mother (*ma*) or father (*pa*) which are generated as *ma*=ceil(*N*\*rand(1, *N*/2)) and *pa*=ceil(*N*\*rand(1, *N*/2)), where ceil rounds the value to the next highest integer i.e. *pops*=4, a random number generator generates following two pairs of random numbers: (0.6710, 0.8125) and (0.7931, 0.3041). Then the following chromosomes are randomly selected for mating: *ma* = [2 3] and *pa* = [3 1]. In this paper, for BGA we use double point crossover, everything between these two points is swapped between the parent chromosomes [17]. The BGA crossover is shown in Figure 3.

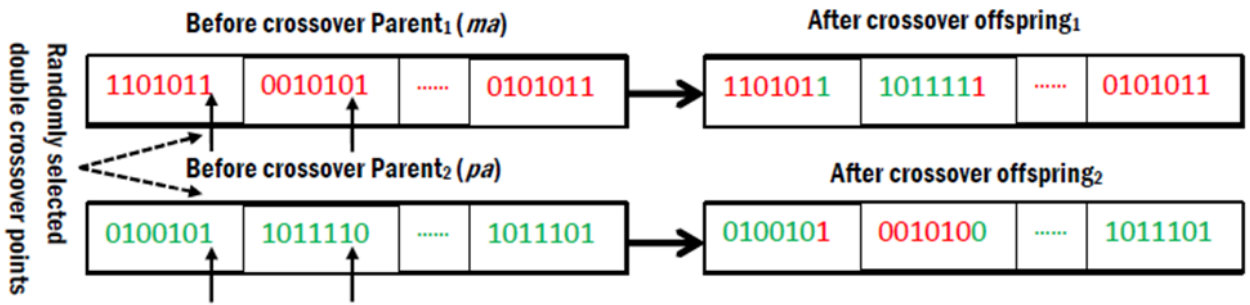

Figure 3. BGA crossover operation

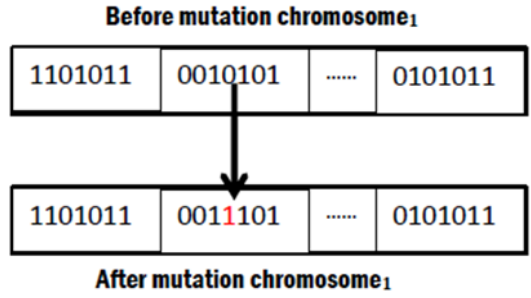

Figure 4. BGA Mutation operation

Next step is mutation operation. For BGA the total no of variables that can be mutated are ≈ (mutation rate* population* size no of bits per variable). The row and column number is nominated randomly and then the nominated bit flips to corresponding binary digit with single chromosome at a time. Figure 4 shows BGA mutation operation. The mutation operation actually helps to provide new search space. In conclusion the proposed GA based optimization algorithm for cooperative spectrum sensing proceeds as follows:

**Step 1:** Set $t=0$ and randomly generate a population of *pops* chromosomes each of which is ($M*nbits$) bits long, where $M$ is the number of secondary users in the network and *nbits* is the number of bits represent each chromosomes.

**Step 2:** Decode each chromosome in the random population into its corresponding weighting coefficients vector Where the weighting coefficients vector $\vec{\omega} = [\omega_1, \omega_2, ..., \omega_M]^T$; $\omega_i \geq 0$ satisfying the condition; $\|\vec{\omega}\| = 1$ which is used to optimize the detection performance.

**Step 3:** Normalize the weighting coefficient vector dividing $\vec{\omega} = [\omega_1, \omega_2, ..., \omega_M]^T$ by its 2-norm such that $\vec{\omega}_i^* = \overline{\omega_i} \Big/ \left( \sum_{i=1}^{M} (\omega_i)^2 \right)^{1/2}$, the constraint $\left\| \vec{\omega}_i^* \right\|_2 = 1$ has to be satisfied.

**Step 4:** Compute the fitness value of every normalized decoded weighting vector, $\vec{\omega}_i^*$ rank their corresponding chromosomes according to their fitness value and identify the best chromosomes $\lfloor pops*elite \rfloor$, where the *elite* is a parameter determines a fraction of *pops* i.e *elite* $\in [0,1)$ and $\lfloor . \rfloor$ denotes floor operation

**Step 5:** Update $t=t+1$ and reproduce $\lceil pops*(1-elite) \rceil$ new chromosomes (candidate solutions) using genetic algorithm operations: selection, crossover and mutation where $\lceil . \rceil$ denotes ceiling operation.

**Step 6:** Construct a new set of population *pops* by concatenating the newly $\lceil pops*(1-elite) \rceil$ reproduced chromosomes with the best $\lfloor pops*elite \rfloor$ found in P(t-1).

**Step 7:** Decode and normalize the chromosomes of the new population *pops* as in **Step 2** and **Step 3** respectively

**Step 8:** Evaluate the fitness value of each chromosome as in **Step 4**

**Step 9:** If t is equal to predefined number of generation(iterations) *ngener,* stop. Otherwise go to **Step 5**

## 5. RESULTS AND DISCUSSION

### 5.1 Testing GA for Optimal Set of Parameters

The GA algorithm has been simulated with different values of same parameter to find out the optimal set of parameters. For simulation the different parameters has been used is mentioned in Table 1.

Table 2: Different GA parameters used for testing

| Parameter name | Rate used |
|---|---|
| Bits per variable *(bits)* | [2,4,6,8,10] |
| Population size (*pops*) | [10, 20, 30,40,50] |
| crossover rate (*Pc*) | [0.50, 0.65, 0.75, 0.85, 0.95] |
| mutation rate ($p_m$) | [0.01, 0.1, 0.15, 0.2, 0.3, 0.6, 0.9] |
| population for reproduction (*Prep*) | [0.5, 0.6, 0.7, 0.8, 0.9] |

And total number of realizations averaged for 100 times. And from simulation results we found the optimal parameters for our CR problem are in Table 2.

Table 3: Optimal set of GA parameters

| Parameter name | Optimal parameter value |
|---|---|
| Bits per variable (bits) | 10 |
| Population size (*pops*) | 50 |
| Crossover rate (*Pc*) | 0.95 |
| Mutation rate ($p_m$) | 0.01 |
| Population for reproduction rate(*Prep*) | 0.9 |

In this paper, a given probability of false alarm $P_f$ = 0.25 has been used for further calculation.

### 5.2 Performance of GA method

In this section, we simulate the proposed GA method with optimal set of parameters which have obtained from the previous section. According to our paper in Figure 3 fitness value defines the probability of detection. It can be seen that proposed GA solution converges and obtain maximum achievable solution that is 1. The parameters is used for BGA are no of generations (*gene*r) = 200, population size (*pops*) =50, secondary users (*M*) = 18, crossover rate ($P_c$) = 0.95, mutation rate pm = 0.01, percentage of population for reproduction Prep =0.9 and probability of false alarm $P_f$ = 0.25.

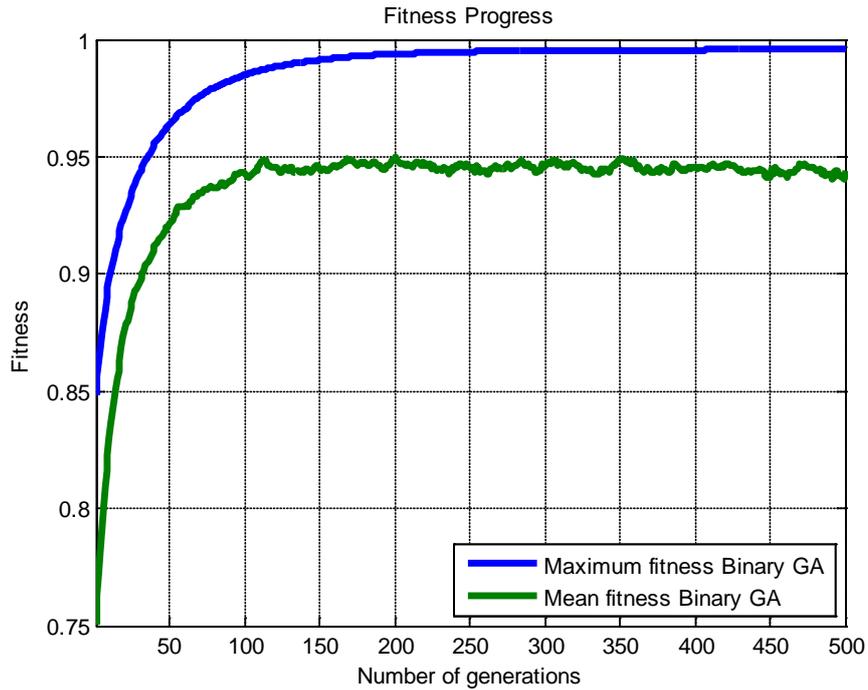

Figure 5: Performance of GA method

**5.3 Performance Comparison of Proposed BGA Method with other conventional method**

The proposed GA is compared with conventional NDC, MDC, MRC and EGC based SDF schemes as well as the OR-rule based hard decision fusion (HDF). The default sensing time and sensed bandwidth are set as $T_s = 25$ us and $B = 6$ MHz, respectively. The relay transmit power is set to 12 dBm and the channel gains of the PU-SU and SU-FC are $\{g_i\}$ and $\{hi\}$, is normally distributed but remain constant within each sensing interval $T_s$, as $T_s$ is sufficiently small. $\{g_i\}$ and $\{h_i\}$, is randomly-generated so that a low SNR environment at SU and FC is realized (SNR < -10 dB). The simulation results are obtained from $10^5$ realizations of channel gains and noise variances.

Figure 4 shows the performance comparison of the proposed BGA-based method versus conventional NDC, MDC, MRC and EGC based SDF schemes as well as the OR-rule based hard decision fusion (HDF). The detection performance is characterized by the ROC curve which is obtained by plotting is the best detection performance comparing to NDC, MDC, MRC and EGC based SDF schemes as well as the OR-rule based HDF.

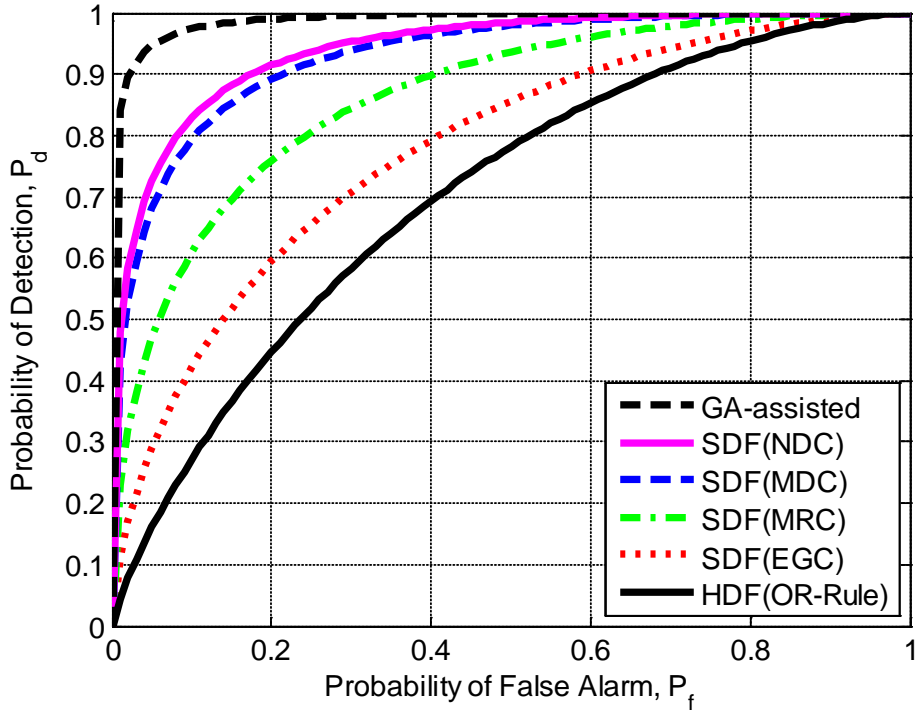

Figure 6: ROC performance comparison of cooperative spectrum sensing using SDF and HDF schemes

The OR-rule scheme, as expected, is inferior to all other methods as it suffers from a significant loss of information content being a HDF process. The EGC SDF scheme shows better performance than the OR-rule HDF scheme but it is inferior to all other SDF schemes due to its fixed and equal weighting coefficients assigned to the energy measurements of the $M$ SUs at the corresponding FC. The MRC-based scheme shows better performance than the EGC one due to its adaptability. The MRC scheme assigns larger weights for the SUs with high SNRs and smaller weights for those with low SNRs and therefore, it controls the contributions of each SU in the overall decision taken at the FC stage. The NDC scheme outperforms the MDC one with non-trivial difference. The detection performance of NDC is slightly better than that of MDC.
It verifies that the computation complexity of the proposed method meets real time requirements of cognitive radio spectrum sensing. The proposed BGA solution converges fast and achieves the optimum performance, which satisfies that BGA-based method is quite stable.

## 6. CONCLUSION AND FUTURE WORKS

A BGA-based method is proposed to optimize the optimal weighting vector required for linear SDF-based at a common fusion centre. The BGA control parameters were first tested and investigated to find the best set suitable for the given CR problem. In this paper, proposed BGA demonstrate as fast and efficient resource allocation algorithms to enable SUs to adapt CRN parameters in the rapidly changing environment. It also verifies that the computation complexity of the proposed method meets real time requirements of cognitive radio spectrum optimization. and it outperforms conventional SDF schemes.

In this paper, Neyman-Pearson Criterion is considered where $P_d$ is maximized for a given $P_f$, and the optimal set of BGA parameters have been found using set-and test approach. In future we will consider continuous genetic algorithm (CGA) and Minimax criterion, where $P_d$ and $P_f$ are jointly optimized, will be taken into account. We will compare BGA and CGA for CRN environment and best algorithm will be used for further research. And then Double-GA engine will be developed where the 1$^{st}$ GA will be used to dynamically find the optimal parameters whereas the 2$^{nd}$ GA will use these optimal parameters to optimize the problem at hand.

**REFERENCES**

[1] Federal Communications Commission, 'Spectrum policy task force report, FCC 02-155', Nov. 2002.

[2] S. Haykin, 'Cognitive radio: Brain-empowered wireless communications', IEEE Journal on Selected Areas in Communications, vo. 23, issue 2, pp. 201–220, 2005.

[3] Ghasemi, A., and Sousa, E.S.: ,'Collaborative spectrum sensing for opportunistic access in fading environments', in Proc. Of IEEE DySPAN 2005, Baltimore, MD, USA, 2005, pp. 131–136

[4] Ganesan and Y. G. Li, "Cooperative Spectrum Sensing in Cognitive Radio, Part I: Two User Networks" IEEE Trans. on Wireless Communications, vo. 6, no. 6, 2007.

[5] Chunmei, G. Q., Wang, J., Shaoqian, L., 'Weighted-clustering Cooperative Spectrum Sensing in Cognitive Radio Context', in Proc. of Int. Conf. on Communications and Mobile Computing, vo. 1, pp. 102-106, 2009.

[6] Z. Chair, P. K.Varshney, 'Optimal data fusion in multiple sensor detection systems', IEEE Trans. Aerospace Electron. Syst. 22 (January) (1986) 98-101.

[7] P. K. Varshney, "Distributed Detection and Data Fusion, Springer," Springer, 1997.

[8] W. Zhang, R. K. Mallik, and K. B. Letaief, 'Cooperative spectrum sensing optimization in cognitive radio networks', in Proc. of IEEE Int. Conf. Communications, pp. 3411–3415, 2008.

[9] Y. C. Liang, Y. Zeng, E. C. Y. Peh, and A. T. Hoang, 'Sensing throughput tradeoff for cognitive radio networks', IEEE Trans. on Wireless Communications, vo. 7, pp. 1326–1337, 2008.

[10] Z. Quan, Shuguang Cui, and Ali H. Sayed, 'Optimal Linear Cooperation for Spectrum Sensing in Cognitive Radio Networks', IEEE Journal of Selected Topics in Signal Processing, vo. 2, no. 1, 2008.

[11] B. Shen and K. S. Kwak, 'Soft Combination Schemes for Cooperative Spectrum Sensing in Cognitive Radio Networks', ETRI Journal, vo. 31, no. 3, 2009

[12] Jun Ma and Ye (Geoffrey) Li, 'Soft Combination and Detection for Cooperative Spectrum Sensing in Cognitive Radio Networks', in Proc. of IEEE Global Communications Conference, pp. 3139-3143, 2007.

[13] Kamran Arshad, Muhammad Ali Imran, and KlausMoessner 'Collaborative SpectrumSensing Optimisation Algorithms for Cognitive Radio Networks', Centre for Communication Systems Research, University of Surrey, Guildford GU2 7XH, UK.

[14] Ayman A. El-Saleh, Mahamod Ismail, Mohd Alaudin Mohd Ali, and Israna H. Arka, 'Hybrid SDF-HDF Cluster-Based Fusion Scheme for Cooperative Spectrum Sensing in

Cognitive Radio Networks', KSII TRANSACTIONS ON INTERNET AND INFORMATION SYSTEMS VOL. 3, NO. 2, December 2011

[15] O. Younis and S. Fahmy, 'Distributed clustering in ad hoc sensor networks: a hybrid energy-efficient approach', Proc. IEEE INFOCOM 2004, pp. 629-640, Hong Kong, China, Mar. 2004.

[16] Ekram Hossain, Vijay Bhargava, 'Cognitive Wireless Communication Networks', Springer, 2007.

[17] Randy L. Haupt and Sue Ellen Haupt, 'Practical Genetic Algorithms', New Jersey: Wiley, 2004.

**Md. Kamal Hossain** completed his B.Eng. degree in Electronics majoring in Telecommunication from Multimedia University, Malaysia, 2012. Currently he is working as Research officer at Multimedia University under Telecom Malaysia R&D grant towards his masters degree. His current research interests include general communication theories, cooperative communication, mobile & satellite communications, cognitive radios and wireless mesh networks.

**Ayman A. El-Saleh** received his B.Eng. degree in Communications from Omar El-Mukhtar University (OMU), Libya, in 1999, his M.Sc. in Microelectronics Engineering from Universiti Kebangsaan Malaysia (UKM), in 2006, and his PhD in Cognitive Wirekess Communications from UKM as well, in 2012. In October 2006, he joined the Faculty of Engineering, Multimedia University (MMU), at which he is currently a Senior Lecturer teaching several telecommunications and electronics courses. He is also a member of ICICE and IACSIT international bodies. His research interests include wireless communications, spectrum sensing techniques, cognitive radio, soft computing using genetic algorithm and particle swarm optimization and FPGA-based digital system design.